# CMS Detector Sensitivity For The Discovery Of SUSY With Two Same-sign Leptons, Jets, And Missing Energy

Ronald Remington[a], on behalf of the CMS Collaboration

[a]*Department of Physics, University of Florida, P.O. Box 118440, Gainesville, FL*

**Abstract.** There is strong theoretical motivation for the study of events with 2 same-sign leptons, jets, and missing transverse energy (MET) at the Large Hadron Collider (LHC). There are many compelling models, for instance, supersymmetry and extra dimensions, which predict such signatures. On the other hand, same-sign lepton production is very suppressed in the Standard Model. Preparations are underway for this search using the Compact Muon Solenoid (CMS) experiment, which will take data at the LHC, scheduled to start running in 2009. The CMS detector has very good measurement capabilities for muons, electrons, jets and MET, and methods are being developed to perform sensitive searches using the first collected data. Using results from a previous Monte Carlo study based on this signature at a center-of-mass energy of 14 TeV, we make discovery potential projections for the early 10 TeV LHC runs. While the results of these extrapolations are encouraging, they are based purely on Monte Carlo and, as such, should be considered only as a rough guide. Data-driven background estimations will be critical for the success of this analysis and are under development.

## THE SIGNATURE

A few fundamental questions concern any search for new physics. For example, what exactly is the signature, and how often is it produced relative to other possible signatures of new physics? How well does the detector measure the final state objects contained in the signature? What are the physics backgrounds for this signature and how often are they produced relative to other signatures?

Qualitatively, the answers to all of these questions provide great motivation for same-sign dileptons as a signature for new physics searches, supersymmetry in particular. The additional requirements of jets and missing energy further enhance the discriminating power of this new physics signature from the Standard Model.

### Production Mechanisms of Same-sign Dileptons

There are many ways to produce two or more isolated leptons at the LHC if SUSY exists:

1) Direct chargino or neutralino production and subsequent decays to leptons
2) Direct squark production and subsequent decays to charginos and neutralinos as in 1)
3) Direct gluino production and subsequent decays to squarks as in 2).

Some of these production mechanisms lead to opposite-sign leptons (OS) and some lead to same-sign leptons (SS). If we assume an mSUGRA scenario with R-Parity conservation, missing energy is guaranteed in all of the three production mechanisms, while jets are guaranteed in the latter two. The OS/SS production ratio is largely dependent on the mass hierarchy between the squark and gluino. For example, if the gluino mass is significantly less than the squark mass, then we would expect nearly equal numbers of OS and SS dilepton events, since gluino-gluino production will dominate and gluinos are blind to charge. Also, the same-sign dilepton final state is a bit more favorable at the LHC than at the Tevatron because it is a proton-proton collider. The valence quarks are slightly more probable to have a same-sign initial state.

While arguments can be made to motivate the SS dilepton signature versus the OS dilepton signature based on possible production mechanisms, the more powerful argument comes from the physics backgrounds. The Standard Model simply does not produce same-sign, isolated dileptons very frequently, while it does produce opposite-sign, isolated dileptons in relative abundance via Drell-Yan and top-antitop ($t\bar{t}$) production. With the exception of diboson production, which is typically a second order weak interaction, there are simply no significant physics processes, which yield SS, isolated dileptons. This is not to say that the CMS detector will not reconstruct SS dilepton events from Standard Model processes [1]. There are spurious sources of SS dilepton events from charge mismeasurement or heavy flavor decays, which need to be considered. Data-driven measurements of these and other types of backgrounds are under active development in CMS.

## CMS Monte Carlo Study at $\sqrt{s}$ = 14 TeV

A study of the discovery potential of the same-sign dimuon channel was performed in 2006 for the CMS Physics Technical Design Report Vol. II [1]. Designed as a counting experiment and optimized for mSUGRA, the study assumed nominal LHC conditions and was based on 10 fb$^{-1}$ of integrated luminosity. The event selection criteria focused on a handful of robust variables with several cut values determined by a genetic algorithm called GARCON [2]. The selection cuts are shown in Table 1 [3].

**TABLE 1.** Selection cuts applied for the same-sign dimuon analysis.

| Trigger | DiMuon HLT | = | "accept" |
|---|---|---|---|
| Quality pre-selection | All Muons $P_T$ | ≥ | 10 GeV/$c$ |
|  | All jets $E_T$ | ≥ | 50 GeV |
| SUSY-distinguishing selection | $\chi^2_\mu$ | ≤ | 3 |
|  | nHits$_\mu$ | ≥ | 13 |
|  | 1$^{st}$ $\mu$ Iso$_\mu$ | ≤ | 10 GeV/$c$ |
|  | 2$^{nd}$ $\mu$ Iso$_\mu$ | ≤ | 6 GeV/$c$ |

|  |  |  |
|---|---|---|
| same sign $n_\mu$ | $\geq$ | 2 |
| nJets | $\geq$ | 3 |
| 1st Jet $E_T$ | $\geq$ | 175 GeV |
| 2nd Jet $E_T$ | $\geq$ | 130 GeV |
| 3rd Jet $E_T$ | $\geq$ | 55 GeV |
| Missing $E_T$ | $\geq$ | 200 GeV |

The performance of these cuts was tested on a small ensemble of points in the mSUGRA parameter space. These points are referred to as the CMS mSUGRA Model Benchmark Points and have the properties listed in Table 2 [3][4].

**TABLE 2.** CMS mSUGRA Benchmark Points.

| Model | $m_0$ [GeV] | $m_{1/2}$ [GeV] | $\tan(\beta)$ | $A_0$ | $sign(\mu)$ | $\sigma_{LO}$ [pb] (14 TeV) | $\sigma_{LO}$ [pb] (10 TeV) |
|---|---|---|---|---|---|---|---|
| LM0 | 200 | 160 | 10 | -400 | + | 240 | 110 |
| LM1 | 60 | 250 | 10 | 0 | + | 41.9 | 16.1 |
| LM2 | 185 | 350 | 35 | 0 | + | 7.4 | 2.4 |
| LM4 | 210 | 285 | 10 | 0 | + | 19 | 11.8 |
| LM5 | 230 | 360 | 10 | 0 | + | 6 | 1.9 |
| LM6 | 85 | 400 | 10 | 0 | + | 4 | 13 |
| LM7 | 3000 | 230 | 10 | 0 | + | 10.2 | 2.9 |
| LM8 | 500 | 300 | 10 | -300 | + | 8.8 | 2.9 |
| LM10 | 3000 | 500 | 10 | 0 | + | 0.178 | 0.06 |

The signal and background yields after applying the event selection criteria in Table 1 are shown in Table 3 [3]. The results of this analysis are scaled to an integrated luminosity of 10 fb$^{-1}$, which is substantially more data than we now expect to accumulate during the 2009-2010 run. The QCD dijet background yields zero events, which could be an artifact of limited statistics as the cross-sections are very high. The significance is also calculated for each signal point.

**TABLE 3.** Signal and background event yields and significance for 10 fb$^{-1}$ at 14 TeV.

|  | QCD | $t\bar{t}$ | W/Z | Diboson | LM1 | LM2 | LM4 | LM5 | LM6 | LM7 | LM8 | LM10 |
|---|---|---|---|---|---|---|---|---|---|---|---|---|
| Events | 0 | 1.5 | 0 | 0 | 341 | 94 | 90 | 61 | 140 | 82 | 294 | 4 |
| Signif | - | - | - | - | >37.0 | 17.6 | 17.2 | 14.0 | 22.3 | 16.3 | 35.9 | 2.2 |

## Projections for 10 TeV with 200 pb$^{-1}$ of Data

Given the encouraging results from the 14 TeV study presented in Table 3, it is worth exploring the prospects of discovery with the data collected during the early era of LHC running at 10 TeV. Initial plans were to take anywhere from 100 pb$^{-1}$ to 200 pb$^{-1}$ of integrated luminosity during the 2009-2010 runs. Since we know the leading order cross-sections at 14 TeV and 10 TeV for the backgrounds and the signal points, we can make projections of how this analysis would perform with the early data. From Table 2 we see that the signal cross-sections are reduced by a factor of roughly three. The $t\bar{t}$ cross-section is reduced by a factor of roughly two. We assume an optimistic scenario with 200 pb$^{-1}$ of integrated luminosity, which corresponds to a reduction by a factor of 50 with respect to the previous study. An important update to this analysis, which is currently in progress, is the inclusion of electron-electron and electron-muon channel. It is expected that these additional final states will increase statistics by as much as a factor of three. Lepton universality would suggest an increase by a factor of four; however, electron reconstruction efficiency is anticipated to be slightly worse than that of muons at CMS. Thus, we have a simple formula to estimate the event yield for a given process,

$$N_{10\text{TeV}} = \frac{3}{50} N_{14\text{TeV}} \frac{\sigma_{10\text{TeV}}}{\sigma_{14\text{TeV}}} \qquad (1)$$

The outcome of this projection can be seen in Table 4. The significance is also recalculated based on these event yields. Systematic uncertainties are not considered in these projections while they are for Table. 3. Benchmark points LM1 and LM8 yield enough events to suggest discovery potential, while LM4 yields enough events to perhaps be excluded. While this scaling exercise makes many assumptions about the efficiencies of the analysis cuts at this lower center-of-mass energy, especially in the case of the $t\bar{t}$ background, the results are encouraging. An argument can be made that the sensitivity to mSUGRA in the same-sign dilepton channel at 10 TeV is not lost, even with a modest amount of data. Certainly, data-driven background estimations will be critical for this analysis and are under development.

TABLE 4. Projected signal and background event yields and significance with 200 pb$^{-1}$ at 10 TeV.

|        | QCD | $t\bar{t}$ | W/Z | Diboson | LM1 | LM2 | LM4 | LM5 | LM6 | LM7 | LM8 | LM10 |
|--------|-----|-----|-----|---------|-----|-----|-----|-----|-----|-----|-----|------|
| Events | 0   | 0.06 | 0   | 0       | 7.9 | 1.8 | 3.4 | 1.2 | 2.7 | 1.4 | 5.8 | 0.1  |
| Signif | -   | -   | -   | -       | 7.7 | 2.7 | 4.3 | 1.8 | 3.7 | 2.2 | 6.3 | 0    |

## ACKNOWLEDGMENTS

I would like to thank all of my colleagues and the faculty members at the University of Florida who helped in the preparation of this report: Didar Dobur, Alexey Drozdetskiy, Andrey Korytov, Konstantine Matchev, Guenakh Mitselmakher, Yuriy Pakhotin, and John Yelton. I would also like to thank: Oliver Buchmueller, Jeff Richman, Frederic Ronga, David Stuart, and the SUSY 2009 Conference Committee and Northeastern University for the opportunity to report on this analysis.